# Towards Steganography Detection Through Network Traffic Visualisation

Wojciech Mazurczyk, Krzysztof Szczypiorski, Bartosz Jankowski
Institute of Telecommunications
Warsaw University of Technology
Warsaw, Poland
e-mail: {wmazurczyk, ksz}@tele.pw.edu.pl, bjankows.pw@gmail.com

*Abstract* — **The paper presents initial step toward new network anomaly detection method that is based on traffic visualisation. The key design principle of the proposed approach is the lack of direct, linear time dependencies for the created network traffic visualisations. The method's feasibility is demonstrated in network steganography environment by presenting *steg-tomography* methodology and developing the dedicated visualisation tool. To authors' best knowledge this is the first utilization of network traffic visualisations for steganalysis purposes.**

## I. INTRODUCTION

Anomalies in network traffic can be caused by malicious actions that might compromise network security. Thus, anomaly detection methods in telecommunication networks focus on finding illegal activities/events, especially those that can be caused by potential attacker/intruder. The device that offers such functionality is called Intrusion Detection/Prevention System (ID/PS) that is a vital network security component allowing the implementation of the assumed security policy. However, currently such solutions for the telecommunication networks are facing problems caused mostly by huge traffic volume that must be monitored: they are not efficient enough. And in ideal situation traffic analysis should be performed in a near real-time manner to prevent potential attacks as soon as possible. This urges to work on effective and efficient solutions for network anomaly detection.

Anomalies are defined as patterns that do not conform to defined and expected established principle. The essence of the anomaly detection is a problem of finding data patters that do not fit the defined set of the expected behaviour. Anomaly detection methods have wide spectrum of applications ([1], [2]) from detection of credit card frauds, software/hardware malfunctions, military applications to detection of the malicious activities in telecommunication networks.

Anomaly detection in telecommunication networks focuses on finding illegal activities/events, especially those that can be caused by potential attacker/intruder. Illegal activity includes: DoS (Denial of Service) or DDoS (Distributed DoS) attacks, network devices port scanning attempts or hacking of the security measures, hostile devices takeovers or other events that can be harmful for security policy defined for a given network. Other sources of anomalies include worms, viruses or trojans. However, another new threat is currently seen as a rising threat to network security: network steganography ([3], [4]). Such methods may be easily utilised as a tool for data exfiltration or to enable network attacks. Detection of network steganography is still in its infancy and does not have satisfactory methods/devices.

The main aim of network steganography is to hide secret data "inside" the carrier i.e. normal transmissions of users. In an ideal situation hidden data exchange cannot be detected by third parties unaware of the steganography usage. The best carrier for secret messages must have two features. Firstly, it should be popular i.e. usage of such carrier should not be considered as an anomaly itself. The more instances of such carriers (steganographically unmodified) are present and utilized in network the better, because they mask using particular carrier to perform hidden communication. Secondly, modification of the carrier related to inserting the steganogram should not be "visible" to the third party not aware of the steganographic procedure. Contrary to typical steganographic methods which utilize digital media (pictures, audio and video files) as a cover for hidden data (steganogram), network steganography utilizes communication protocols' control elements and their basic intrinsic functionality. As a result, such methods are harder to detect and eliminate.

In the last decade one can observe very intensive research effort related to steganography and its detection methods and to network steganography in particular. This has been caused by two facts: first, industry and business interest in DRM (Digital Rights Management) and second alleged utilisation of the steganographic methods by terrorist while planning attack on USA on September 11[th], 2001 [6]. Incoming years proved that it was not an isolated case: in 2010 it was reported that the uncovered Russian spy ring used digital picture steganography to leak classified information from USA to Moscow [7].

In order to minimize the potential threat to public security, identification of such methods is important as is the development of effective detection (steganalysis) methods. As mentioned above, up till now no universal, widely deployed detection methods exist that are efficiently able to fight network steganography usage. Therefore, it is important to include in the detection methods development the ability to uncover steganographic communication as well.

That is why, in this paper we make an initial step towards more general steganalysis method based on network traffic visualisation that we named *steg-tomography*. The approach is demonstrated on example of LACK (Lost Audio Packets Steganography) steganographic method that is one of the state of the art steganographic method for VoIP (Voice over IP) and was introduced in 2008 [26]. This solution takes advantage of the fact that in typical multimedia communication protocols, excessively delayed packets are not used for the reconstruction of transmitted data at the receiver; that is, the packets are considered useless and discarded. Thus, for some of the chosen voice packets intentional delay is introduced and their payload is substituted with secret data.

The rest of the paper is structured as follows. Section 2 describes background and related work in anomaly detection, steganalysis and network traffic visualisation for security purposes. Section 3 describes experimental methodology: details of the proposed steganalysis approach, chosen steganographic method and developed testbed. Section 4 presents and discusses obtained experimental results. Finally, Section 5 summarizes our efforts.

## II. RELATED WORK

Anomaly detection can be classified, based on availability of the training data that is used to "teach" detection methods of what the anomalies are (both for normal as well as anomaly class), as [1]:

- **Supervised Anomaly Detection** (examples: [12], [13]) – where there is availability of a training data set (both normal and anomalous). Based on this data proper models are derived. Next, unseen, inspected data is classified based on comparison with these models.
- **Semi-supervised Anomaly Detection** (examples: [14], [15]) – where the training data has instances for only the normal class (which is more desirable). That is why these methods are more widely applicable than supervised techniques.
- **Unsupervised Anomaly Detection** – where training data is not required (which is most desirable), thus such methods are most widely applicable. These methods make the implicit assumption that normal behaviour is far more frequent than anomalies and utilises this rule to detect such defined anomalies.

As mentioned in Section 1, in telecommunication networks the devices that are responsible for anomaly detection are called ID/PSs. ID/PSs usually utilise detection methods from semi-supervised or unsupervised groups. Denning in [16] divided these systems into: host-based intrusion detection systems (HIDS – an example of such IDS/IPS is described in [17]) are dedicated to protect single user host and network-based intrusion detection systems (NIDS – an example of such solution is described in [18]) are used for the whole network protection. Moreover, in ideal situation traffic analysis and anomaly detection should be performed in a near real-time manner to prevent potential attacks as soon as possible. However, current ID/PS systems are, generally, unable to achieve this goal, because they are facing the following problems related to:

- **Huge traffic volume that should be monitored** – many IDS/IPS systems are just not efficient enough to process it in near real-time manner.
- **The difference between normal/expected and anomalous behaviour** that can be sometimes hard to define. For example, network steganography methods intentionally imitate users' network traffic behaviour in order to hide their existence, thus it can be qualified as a normal behaviour. On the other hand real user's traffic can be classified as anomaly in certain circumstances (e.g. implementation of the new protocol or service that was not encountered before by ID/PS).
- **Potential adaptability of the attacker behaviour** in order to omit disclosure.
- **Nature of the telecommunication networks and their services** – normal/expected behaviour can change in time. Thus, expected behaviour in a given time moment does not guarantee successful anomaly detection in the future.
- **Existence of the certain behaviours that are caused by users without negative intentions**. Such behaviours (e.g. caused by device malfunction) cannot be treated as expected/normal behaviour but can make detection of the real network anomalies harder.

Moreover, it must be emphasised, that although ID/PSs are usually including detection of various kinds of anomalies – mainly network attacks, they are not able in their current form to prevent network steganography utilisation.

Network steganography detection methods were somewhat developed independently from ID/PS systems. Most steganalysis methods focus on trying to detect the presence of hidden communication and then on limiting its transmission capabilities, because as stated in [5] and [19] elimination of all network steganography opportunities is practically impossible. Currently, most steganalysis methods can be characterised by high computational complexity and are time consuming, which in practise limits their applicability for real-network traffic. Steganography detection methods can be divided into [20]:

- **Statistical steganalysis** – examples of such solutions include methods based on linear regression SVM (Support Vector Machines) or information theory. Statistical steganalysis methods can rely on some simple rules which are created based on the probability of occurrence of certain events e.g. filling with data some rarely used or optional fields from protocol headers. The example of such method for IPv4 protocol was proposed by Murdoch et al. [21]. Sohn et al. [22] developed solution for detection of steganographic methods that utilise IP Identification (IPv4) and ISN (TCP) fields and is based on SVMs.
- **CI (Computational Intelligence) based steganalysis** – such methods utilise: neural networks, fuzzy logics or genetic algorithms. An example of such methods was proposed by Tumoian et al. [23]. Steganalysis of DSSS method that uses genetic algorithm was developed by Sedghi et al. [24].
- **Hybrid solutions** that incorporates functionality of both of above groups. An example of hybrid steganalysis was introduced by Knapik et al. [25] and it combines utilisation of SVMs and genetic algorithm.

In this paper an initial step towards new anomaly detection method based on network traffic visualisation is presented. To prove the method's feasibility it will be applied in network steganography environment.

Currently, visualization systems are widely used for network security and together they are described by term *security visualization*. Generally, network security benefits from visualization of traffic by facilitating hosts/servers monitoring, analysis of external-internal communications, port activity or routing behaviour. It is also suitable to discover attack patterns (a good survey by Shiravi et al. is available [27]). However, to authors' best knowledge this is the first approach that utilises network traffic visualisation for steganalysis purposes. The method's details and its utilization for network steganography will be explained in the following sections.

### III. Experimental Methodology

#### A. Detection Methodology

To authors' best knowledge this is the first approach that utilises network traffic visualisation for steganalysis purposes. The key design principle of the proposed anomaly detection approach is the lack of direct, linear time dependencies for the created network traffic visualisations.

The detection method functions by observing snapshots (traffic visualisations) of captured network traffic (three chosen parameters) in the defined time frame (window). The time window size depends on many factors like the type of inspected steganographic method, the type of traffic affected, the available resources etc. Of course, the size of the observation window can be extended to e.g. single connection (as it presented in this paper) or the visualisations can be periodically updated while the connection lasts (Fig.1).

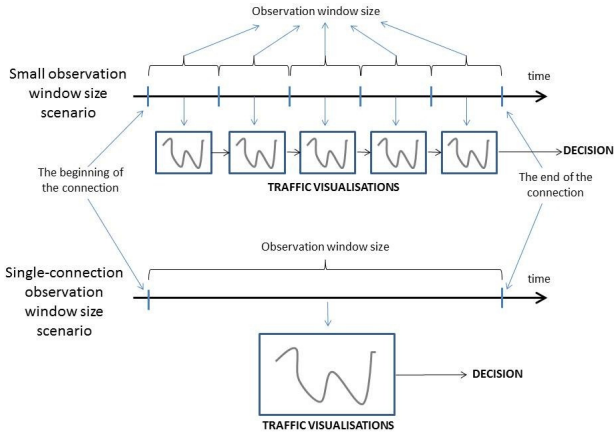

**Figure 1. Detection methodology possibilities**

The approach utilized in this paper for steganalysis purposes that we named *steg-tomography* can be divided into the following steps:
1. Analyse the steganographic method functioning and choose the list of potential parameters for the observation.
2. Perform initial visualisations on steganographic and non-steganographic data to analyse and limit the parameters to the three most promising ones.
3. Analyse and choose the best size of the observation window i.e. the period for which network traffic visualisation(s) will be created.
4. For traffic that needs evaluation preform visualisations according to the selected observation window and after analysis decide whether it carries secret data or not.

#### B. Chosen Steganographic Method and Experimental Setup

LACK is a steganographic method intended for a broad class of multimedia real-time applications like IP telephony or videoconferencing. It was originally proposed in 2008 [26] and is considered as one of the states of the art VoIP steganographic techniques [28].

LACK benefits from the fact that in typical real-time multimedia environments excessively delayed packets are not used for the reconstruction of transmitted data at the receiver, i.e. the packets are considered useless and discarded. It impacts the interior of the packets that carry user's voice (RTP packets) as well as their time dependencies.

The overview of LACK functioning is demonstrated in Fig. 2. At the transmitter (Alice), one voice packet is selected from the voice stream and its payload is substituted with bits of the secret message (1). Then, the selected RTP packet is intentionally delayed prior to its transmission (2). Whenever an excessively delayed packet reaches a receiver unaware of the steganographic procedure, it is discarded, because it interprets the hidden data as "invisible". However, if the receiver (Bob) is aware of the hidden communication, then, instead of dropping the delayed RTP packet, it extracts the hidden data from payload (3).

The detection of LACK is difficult for the following reasons:
- Delayed RTP packets can be treated as a natural phenomenon for IP telephony, thus introducing intentionally delayed voice packets cannot be considered as anomalous behavior per se.
- The size of receiving (jitter) buffer is not known in advance for third parties because it can be of a various fixed value or can be adapted dynamically e.g. to network conditions. Thus, from the steganalysis point of view it is hard to assess whether delayed RTP packets will be really utilized for voice reconstruction at the receiving end or will be treated as useless and discarded.

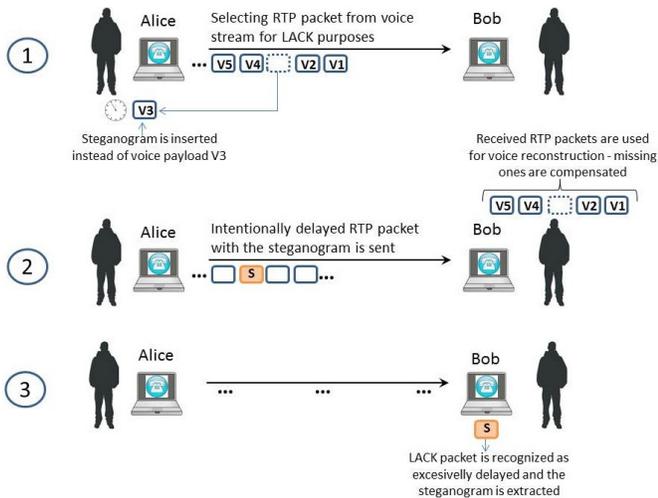

Figure 2. The idea of LACK

Considering above it is important to develop effective LACK detection method.

In this paper we will try to detect and evaluate LACK implementation called StegoSIP [29] that was developed as a part of thesis project under supervision of Prof. Luigi Ciminiera of Politecnico di Torino and is available under GNU GPLv3 licence. The application allows to control two of most the important LACK parameters i.e. the *delay* of the chosen voice packets and the *frequency* of steganographic packets.

Based on the LACK operation analysis (Step 1 in Sec. IIIA) the following six parameters for inspected RTP stream were selected for further visualisation analysis:

- Payload
- Sequence number
- The size of the packet
- Payload Type
- SSRC/CSRC identifiers
- Jitter

The experimental testbed for LACK detection is based on two virtual machines running on VMware Server. Both machines are running Linux Debian 6.0 Squeeze x86 and are installed with free VoIP softphone Ekiga [30] and is configured with defaults: Theora for video and Speex for speech coding. StegoSIP is applied to the Ekiga's traffic. Hosts establish VoIP connections in controlled LAN (Local Area Network) environment and the resulting traffic is captured using Wireshark sniffer [31] and further processed with the help of developed, specialized 3D visualisation tool.

## C. Dedicated 3D Visualisation Tool

Captured network traffic visualisation was performed with the use of the dedicated visualisation tool named Easy 3D Plotter. It was developed in IDE Matlab 7.11 (R2010b) environment. For effective plotting *plot3k* method [32] was modified and utilized. The GUI of the tool is presented in Fig. 3.

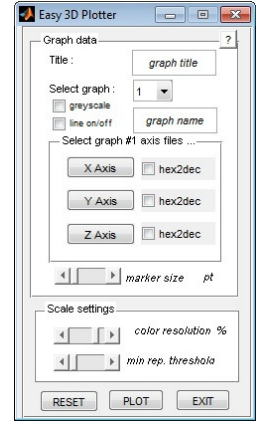

Figure 3. The GUI of the developed visualisation tool

The tool allows to create up to ten 3D diagrams that can be customised independently. The buttons in *axis files…* section allows to link each of the figure's dimension to the file with captured data. The rest of the buttons influence the form of the resulting diagram and due to limited space will be not provided here.

## IV. EXPERIMENTAL RESULTS

### A. Initial LACK analysis and results

For the experiments StegoSIP implementation was configured with advised defaults i.e. the delay of the chosen voice packets was set to 0.5 s and the frequency of steganographic packets to 5 i.e. 1 for every 5 packets is utilized for steganographic purposes.

The initial visualisation analysis for exemplary VoIP connection with LACK applied is presented in Fig. 4.

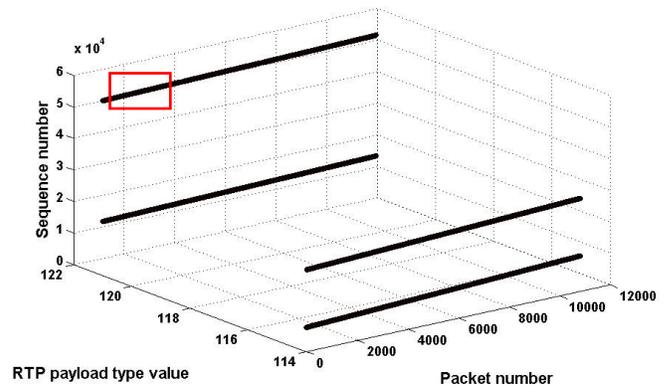

Figure 4. Exemplary steganographic VoIP call analysis using visualisation tool

The analysis was performed considering the order of the RTP packets that enter receiving network interface, their sequence numbers and Payload Type field. It can be seen that single Ekiga connection consists of four distinct RTP streams with payload type of 121 (Theora) and 114 (Speex). In shown diagram it looks like there is no sign of LACK-related anomalies. However, if we zoom in a fragment (red square in Fig. 4) of one of RTP streams one can see the discontinuities (Fig. 5). Of course lost or reordered RTP packets can be treated as a natural phenomenon that happens often in IP networks. However, as experimental results presented in [8] indicated packets' reordering is rather rare. Moreover, packets in all RTP streams should be affected which is not the case here, thus, this can be treated as anomalous behaviour.

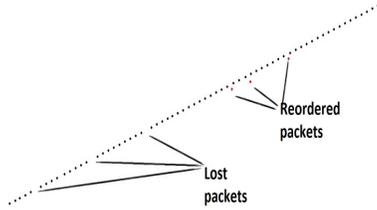

**Figure 5. Discontinuities in RTP stream**

This effect is also visible when we consider subsequent voice packets sequence numbers' difference (see Fig. 6). In fact, the change in the RTP stream packets' sequence causes two anomalies. First, we see the delayed packets that are marked red. Second, the gap caused by these packets is also visible in the place where they originally were (black points in the blue ellipse).

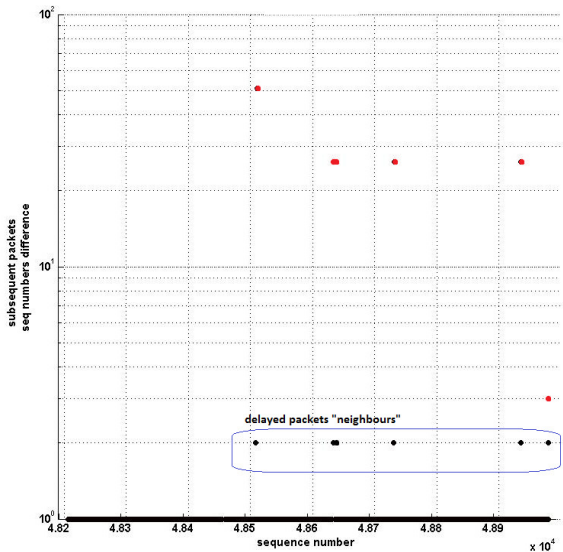

**Figure 6. Subsequent voice packets sequence numbers' difference (logarithmic scale) for a fragment of sequence numbers**

Results presented above prove only that there is an anomaly (delayed RTP packets of payload type equal to 121) but we cannot unambiguously state that the hidden communication is taking place.

Now let us consider the size of the RTP packets (Fig. 7) together with their sequence numbers and the order they enter receiving network interface. As can be seen the delayed packets (marked with red) were of a size larger than 100B.

Next, let us analyse the RTP packets' payload. In Fig. 8 X and Y axes present 4 bytes of payload (oldest and random) and Z axis describes whether the packet had subsequent sequence number (0) or not (1).

Careful analysis of Fig. 8 shows that some payload bytes can be observed repeatedly (see white and yellow points marked in green ellipse that repeated from 2 to about 10 times). However, it must be noted that single and fixed payload value was observed more than 50 times (marked in red) and it was different for delayed and not delayed RTP packets. Thus, it can be concluded that in the delayed voice packet the steganogram is carried and it is not ciphered. It turns out that StegoSIP implements the whole TCP/IP stack inside the steganographic tunnel created with use of LACK. That is why, only packets of a large size are chosen. However, this in turn makes LACK more susceptible for steganalysis.

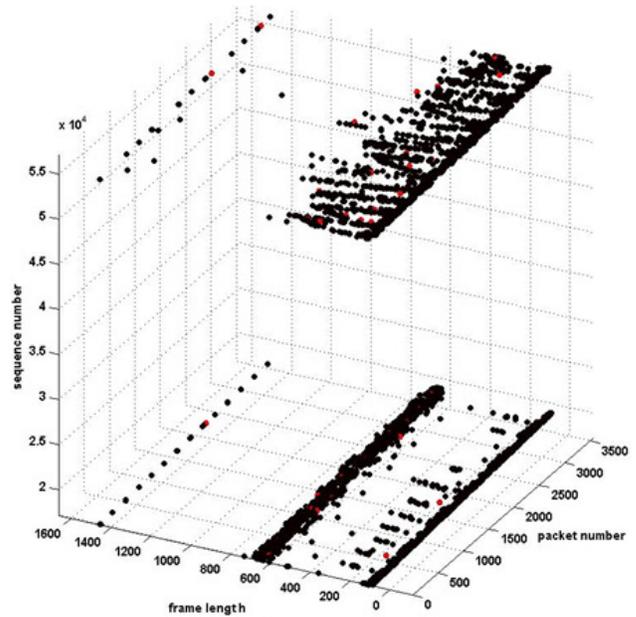

**Figure 7. RTP packets size analysis**

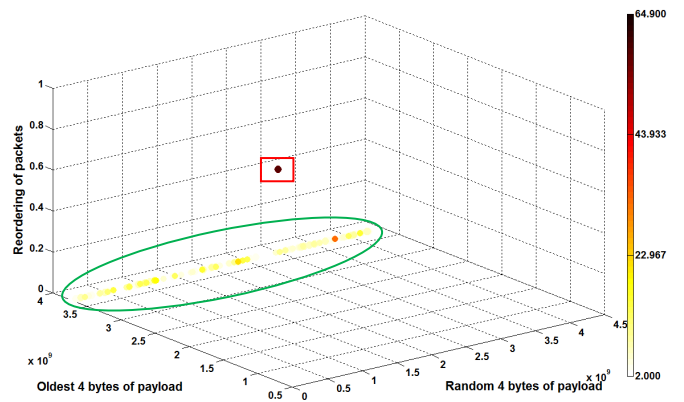

**Figure 8. Anomalies caused by StegoSIP**

Parts of the protocols' headers that are carried inside the steganographic tunnel are of a fixed size and values that can be easily detected. Moreover, they introduce the significant overhead on LACK steganographic bandwidth.

### B. StegoSIP detection

Let us now compare non-steganographic and steganographic VoIP connection. Based on the results provided in previous subsection in order to detect this LACK implementation it is advised to observe:

- **Sequence numbers or jitter** of the RTP packets to be able to verify whether the inspected packet was delayed and/or is out of order.
- **Fragments of payload of RTP packets** – analy can be helpful to check if any other type of paylo besides voice or video is carried.
- **RTP packets' size** – relationship between delay the packets and their size can be inspected.

Considering above, the comparison of steganographic a non-steganographic VoIP connection is presented in Fig. The following three parameters are observed: difference sequence numbers of the consecutive RTP packets, fra length and first 4 bytes of payload. The results presented Fig. 9 are limited to only single RTP stream. Inside t covert channel SSHv2 (Secure Shell v2) session is tunnelle In result, the most of the steganographic payload is ciphe using AES (Advanced Encryption Standard). For present experiments below G.711 codec was utilized for vo connections. The StegoSIP configuration is default and the same as in the previous subsection.

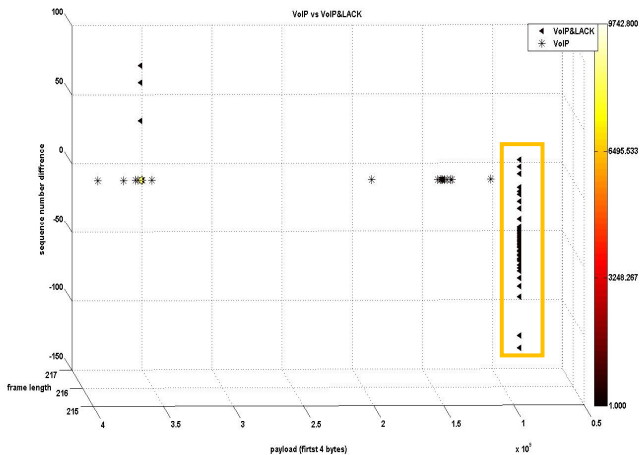

**Figure 9. Comparison of steganographic and non-steganographic VoIP connection**

Fig. 9 shows that all packets were of the fixed size (217 bytes) which is a result of the utilized voice codec. RTP packets for typical, non-steganographic VoIP call are aligned. For VoIP steganographic connection a lot of RTP packets were out of sequence and additionally they experienced the same 4 bytes of the payload (yellow square). This proves that the LACK steganographic method was applied to VoIP call and thus it can be easily detected.

In the next step we want to observe if StegoSIP will be still visible if the frequency of steganographic packets is decreased from 5 to 20 i.e. now 1 for every 20 packets is utilized for steganographic purposes.

For non-steganographic VoIP call (Fig. 10) the same parts of payload repeated more than 7000 times and for steganographic one only 4000 times (Fig. 11). Moreover, in the latter figure we can observe a lot of out-of-sequence voice packets for the same payload value. This proves that even decrease in steganographic bandwidth still makes StegoSIP detectable.

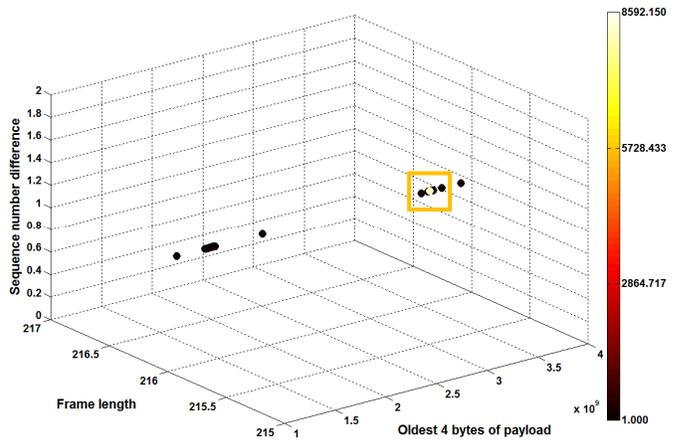

**Figure 10. Non-steganograhic VoIP call**

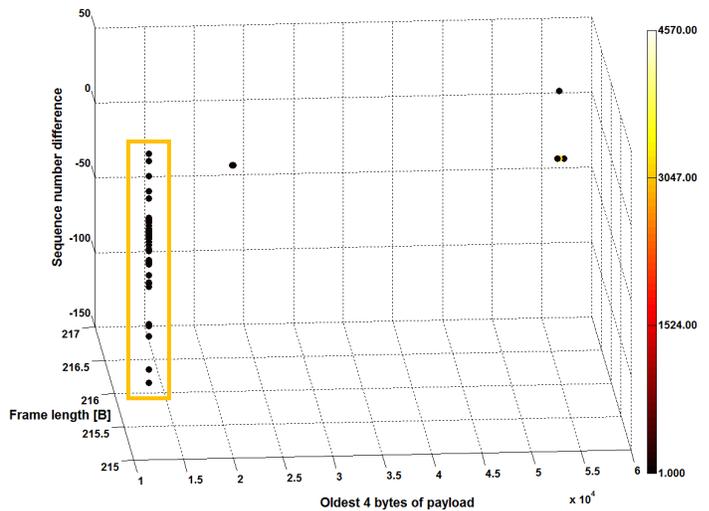

**Figure 11. VoIP call with LACK applied**

### C. Discussion

StegoSIP implementation of LACK can be fairly easy detectable especially after taking into consideration results

from Section IVA. This is possible for the following reasons:
- StegoSIP does not cipher the secret data that is inserted into excessively delayed RTP packets.
- StegoSIP tunnels the whole TCP/IP stack inside the covert channel and thus it makes some payload bytes repeatable (network protocols' fields).
- StegoSIP impacts RTP stream by introducing discontinuities in sequence numbers that as it was previously researched by Mazurczyk et al. [8] is rather rare for VoIP. The better solution is to delay the whole stream to invoke receiver's buffer overflow or underflow.
- StegoSIP delays packets by a fixed and always the same configurable value, which makes steganalysis easier.
- StegoSIP tunnels secret data only in the large-sized packets.

For these reasons detection of StegoSIP is possible even if the LACK steganographic bandwidth is decreased. However, if the abovementioned issues are taken into account and the implementation is upgraded then similar analysis should be carried out to once again assess its detectability.

## V. CONCLUSIONS

In this paper the first utilization of network traffic visualisations for steganalysis purposes was carried out. The key design principle of the proposed approach is the lack of direct, linear time dependencies for the created network traffic visualisations. This analysis can be treated as an initial step towards new anomaly detection method. Of course this paper had not revealed all of the benefits that network traffic visualisation provides for steganalysis. There is still a lot of future work that should be considered:
- Different sizes of the observation window in which the snapshots of network traffic are performed can be researched. This can provide opportunity of faster detection.
- Propose an algorithm that will automatically analyse consecutive visualisations for evaluated traffic/connection e.g. based on clustering.
- The effectiveness of the proposed detection method must be compared with some of the most popular steganalysis ones.

Of course, if the experiments are repeated in the real IP network e.g. in the Internet such analysis can be harder to perform. It must be emphasised that StegoSIP implementation issues i.e. periodic choice of steganographic packets and fixed delays would be visible also in real-life networks. However, if these implementation issues will be resolved then additional experiments should be performed to verify its detectability. This is considered as a future work.

Moreover, for other steganographic methods that are not trivial to detect network traffic visualisation analysis should be developed. This will give opportunity to find some common parameters that should be inspected in order to create more universal detection method that will be successful in disclosing network steganography.

ACKNOWLEDGMENT

This work was supported by the Polish Ministry of Science and Higher Education and Polish National Science Centre under grants: 0349/IP2/2011/71 and 2011/01/D/ST7/05054.